\documentclass[aps,prb,reprint,superscriptaddress,nobalancelastpage,twocolumn,showpacs]{revtex4-1}
\usepackage{color}
\usepackage{comment}
\usepackage{graphicx}
\usepackage{epstopdf}
\usepackage{amsmath}
\usepackage{latexsym}
\usepackage{epstopdf}
\usepackage{amsmath,amssymb}
\usepackage{amssymb,amsmath}
\usepackage{multirow}
\usepackage{mathtools}
\usepackage[dvipsnames]{xcolor}
\newcommand{\beq}{\begin{equation}}
\newcommand{\eeq}{\end{equation}}

\usepackage{tikz}
\usepackage{xcolor}
\usetikzlibrary{patterns}

\begin{document}

\title{Area law and universality in the statistics of subsystem energy}

\author{Khadijeh Najafi }
\affiliation{Department of Physics, Virginia Tech, Blacksburg, VA 24061, U.S.A}
\affiliation{Department of Physics, Georgetown University, 37th and O Sts.  NW, Washington, DC 20057, USA}
\author{ M.~A.~Rajabpour}
\affiliation{Instituto de F\'isica, Universidade Federal Fluminense, Av. Gal. Milton Tavares de Souza s/n, Gragoat\'a, 24210-346, Niter\'oi, RJ, Brazil}

\date{\today{}}

\date{\today{}}

\begin{abstract}

We introduce R\'enyi entropy of a subsystem energy as a natural quantity which closely mimics the behavior of the entanglement entropy and can be defined 
for all quantum many body systems. In other words, consider a quantum chain in its ground state and then, take a subdomain of this system with natural
truncated Hamiltonian. Since the total Hamiltonian does not commute with
the truncated Hamiltonian, the subsystem can be in one of its eigenenergies with different probabilities. Using the fact that the global energy eigenstates are
locally close to diagonal in the local energy
eigenbasis, we argue that the R\'enyi entropy of these probabilities follows an area law for the gapped systems. When the system is at the critical point,
the R\'enyi entropy 
follows a logarithmic 
behavior with a universal coefficient. Consequently, our quantity not only detects the phase transition but also determines the universality class of the critical
point.  Moreover we show that the largest defined probabilities are very close to the biggest Schmidt coefficients. We quantify this by defining
a {\it{truncated}} Shannon entropy
which its value is almost indistinguishable from the {\it{truncated}} von Neumann entanglement entropy.
Compare to the entanglement entropy, our quantity has the advantage of being associated to a natural observable (subsystem Hamiltonian) that can be defined (and probably measured) 
easily for all the short-range
interacting systems.
%Our results show that the Shannon(R\'enyi) entropy of the subsystem energies closely mimics the behavior of the entanglement entropy in quantum chains.
We support our arguments by detailed numerical calculations performed on the transverse field XY-chain.
\end{abstract}
%\pacs{}
\maketitle

%%%%%%%%%%%%%%%%%%%
\section{Introduction}
Quantum information theory and particularly the concept of entanglement has played an increasingly important role in condensed matter and high energy physics
in recent years.
More specifically, it has been very useful in the study of different phases of many body quantum systems 
\cite{Amico2008,Affleck2009,Moore2009,Peschel2009,Cramers2010,Modi2012,Laflorencie2016,Chiara2017}, structure of quantum field theories \cite{CC2009,Doyon2009,Casini2009}
and understanding the gravity\cite{Takayanagibook,Nishioka2018}. Most of these studies are directly related to the concept of entropy which can be defined when
there is a probability distribution. The Shannon entropy
and its natural generalizations, the R\'enyi entropies, are the building blocks of the classical information theory. They can be also 
used as the starting point to define and understand the quantum version of the entropy. The R\'enyi entropy for $\alpha\geq0$ is defined as
\begin{equation}\label{Renyi}
H_{\alpha}=\frac{1}{1-\alpha}\ln\sum_{i=1}^np_i^{\alpha},
\end{equation}
where $p_i$'s are the corresponding probabilities. For $\alpha=1$, one can recover the Shannon entropy $H_1=-\sum_{i=1}^np_i\ln p_i$. 
To see how the above quantities can be used to study the quantum many-body systems and especially quantum phase transitions,
consider a quantum system in its ground state $|\psi_g\rangle$. Since in quantum mechanics we have normally infinite
number of observables we have the freedom to write the ground state in the basis of one of them as $|\psi_g\rangle=\sum_Ia_I|I\rangle$, where $|I\rangle$ 
is one of the eigenstates of the considered observable and $p_I=|a_I|^2$ is the probability of finding the corresponding 
eigenvalue. These probabilities can be used to define the Shannon entropy and one of the immediate consequences is the 
entropic uncertainty relation \cite{uncertainty}. By choosing an appropriate local observable, for example, spin,
one can use the Shannon entropy to detect the phase transitions and restricted versions of universalities \cite{estephan2010a,estaphan2010b,Fradkin2007,Oshikawa2017}.
The situation is even more interesting when one considers the marginal probabilities in a subsystem. When the corresponding
observable is a local quantity, the associated probabilities are called formation probabilities and it is known that especial type of them can be used to 
 determine the central charge and the universality class of the critical points \cite{Estephan2014,Rajabpour2015,Najafi2016,Rajabpour2016}. For local observables,
 it is natural to expect that the Shannon(R\'enyi) entropy be proportional with the size of the subsystem (volume-law)\cite{footnotevolumelaw} which makes the leading term
 non-universal and less interesting. However, numerous numerical calculations suggest that for critical quantum chains, the subleading term
 follows a logarithmic behavior with respect to the subsystem size with a coefficient which is universal and connected to the central charge \cite{AR2013,estephan2014b,AR2014,AR2015,Najafi2016,Alcaraz2016}.
 For other related studies see Refs.\cite{Luitz2014a,Luitz2014b,Luitz2015c}.  Although from the experimental point of view,
 the Shannon (R\'enyi) entropy of local observables in a subsystem looks a natural
 quantity,  theoretically, it is interesting to make the quantity "basis independent" by minimizing over all the possible bases. This minimization leads to the von Neumann 
 entanglement entropy and its generalizations quantum R\'enyi entropies\cite{WildeBook}. For a density matrix $\rho$,
 the quantum R\'enyi entropy for $\alpha\geq0$ is
\begin{equation}\label{quantum Renyi}
S_{\alpha}=\frac{1}{1-\alpha}\ln\text{tr} \rho^{\alpha}.
\end{equation}
For $\alpha=1$, we recover the von Neumann entropy $S=-\text{tr}\rho\ln\rho$. If we consider the total system be in a pure state (for example, the ground state)
and calculate the reduced density matrix for a subsystem and plug this  matrix in the above equation, we end up with the entanglement entropy 
of the subsystem with respect to its complement. Another way of looking to this quantity is by calculating the Shannon entropy in the Schmidt basis, which is a complicated
non-local basis that minimizes the entropy.
von Neumann entropy has been studied in a myriad of articles, 
for review see Refs. \cite{Amico2008,CC2009,Doyon2009,Affleck2009,Moore2009,Peschel2009,Casini2009,Cramers2010,Modi2012,Laflorencie2016,Takayanagibook,Chiara2017,Nishioka2018}. 
The most relevant results for our purpose are the followings: the entanglement entropy of the ground state for gapped systems follows
an area law \cite{Hastings2008,Cramers2010,Brandao2013,Vazirani2012,Brandao2015}. For infinitely long one dimensional critical systems the R\'enyi (von Neumann) entropy of a subsystem with size
$l$ is given by\cite{Holzhy1994,Vidal2003,CC2004}
\begin{equation}\label{Renyi conformal}
S_{\alpha}=\frac{c}{6}(1+\frac{1}{\alpha})\ln l+\gamma_{\alpha},
\end{equation}
where $c$ is the central charge of the underlying conformal field theory and $\gamma_{\alpha}$ is a non-universal constant. Based on the above results,
one can make the following argument: Shannon entropy of a subsystem for local observables (basis) follows a volume law but
it follows an area law (logarithmic law) in gapped (critical) systems for non-local Schmidt basis. 
This makes one to believe that there should be some non-local observables in between these two extreme cases. For example, one can think about
the total number of particles in a subsystem and study its distribution as it has been done in the context of full counting statistics 
\cite{Eisler2003,Klich2006,Klich2009,Calabrese2012,Rachel2012a,Rachel2012b,Klich2014,FCS}. One can also study other quantities such as the  total magnetization
distribution \cite{Fendley2008} or the distribution of the subsystem energy \cite{Najafi2017}. 

Among the many natural non-local 
quantities that one can study, we are interested in the one which, apart from being natural,  can be defined for all the quantum chains and 
can mimic in the best way the entanglement entropy of a subsystem.
 Since the von Neumann entropy is defined without reference to the observable of the system, it is widely believed that the direct measurement of this quantity is impossible due to the fact that this quantity is nonlocal and it's measurement requires knowledge about the full reduced density matrix which grows exponentially with the size of the system. An observable such as the truncated Hamiltonian that its statistics closely mimics the entanglement Hamiltonian is not only interesting by itself, might also be useful in a better theoretical and experimental understanding of the entanglement entropy. 

%{\color{red} Since the von Neumann entropy is defined without reference to the observable of the system, the direct %measurement of this quantity without full tomography of the state of the system seems impossible. An observable such as %the truncated Hamiltonian that its statistics closely mimics the entanglement Hamiltonian is not only interesting by %itself, might also be useful in a better theoretical and experimental understanding of the entanglement entropy. }

In this paper, we argue that the subsystem energy is such a quantity. Its R\'enyi entropy follows an area law for the gapped systems and 
it is logarithmic for the critical systems. The coefficient of the logarithm is universal and may well be related to the central charge. 
Compare to the entanglement entropy the Shannon(R\'enyi) entropy of a subsystem energy has the advantage that it is related to  a 
natural observable which not only can be defined easily but also measured simply for the short-range
interacting systems. 
The paper is organized as follows: in sec. II, based on some known results we first argue that the statistics of the 
subsystem energy is a perfect candidate which can mimic the entanglement entropy. Then, in sec III, we 
develop an elegant method to calculate the distribution of the eigenvalues of the quadratic observables 
in the free fermionic systems  and provide  an exact formula for the probabilities in the basis of the subsystem energy. Then, 
in section IV, we study the R\'enyi entropy of the subsystem energy for the transverse field XY chain 
numerically to demonstrate the validity of the arguments. We further provide our results to confirm the 
R\'enyi entropy of subsystem energy with the R\'enyi entanglement entropy. Finally, in section V, we comment on 
the possible experimental setups that can be used to measure the subsystem energy Shannon entropy.

%%%%%%%%%%%%%%%%%%%
\section{subsystem energy basis}
%\label{sec2}
%%%%%%%%%%%%%%%%%%%
We start by considering a generic nearest neighbor Hamiltonian of an infinite system\cite{footnote1} $\textbf{H}=\sum_{i=-\infty}^{\infty}H_{i,i+1}$, where $H_{i,i+1}$ has support in the set of
sites $i$
and $i+1$. Now, consider $l$ contiguous sites and define
a truncated Hamiltonian for the subsystem as $\textbf{H}_l=\sum_{i=1}^{l-1}H_{i,i+1}$. For interacting Hamiltonians, we always have $[\textbf{H},\textbf{H}_l]\neq0$, which means that 
if the total system is in its ground state, the subsystem can be in different eigenstates $|e_j\rangle$ with different probabilities $p(e_j)$. Note that for the short-range Hamiltonians 
the right-hand side of $[\textbf{H},\textbf{H}_l]$ is dependent just on the boundary terms which hints on "weak" uncertainty relations.
To study the corresponding probabilities, one can 
first calculate the reduced density matrix of the subsystem $\rho_l=\text{tr}_{\bar{l}}|\psi_g\rangle\langle\psi_g|$ where the trace is over the complement of the subsystem.
The reduced density matrix is exactly diagonal in the Schmidt basis which leads to the quantum entanglement entropy, but it has off-diagonal terms in every other 
basis. However, an interesting theorem is proved in Ref. \cite{muller2015}, see also Ref.\cite{Arad2016}; which indicates  that global energy eigenstates are locally close to diagonal in the local energy
eigenbasis\cite{footnote2}. In other words, the reduced density matrix is weekly diagonal\cite{footnote2} in the eigenbasis 
of the truncated Hamiltonian. This theorem
suggests that the eigenbasis of the subsystem Hamiltonian is not that much different from the Schmidt basis. 
It is equivalent to say that $p(e_j)$'s are close to the Schmidt coefficients. One can then guess that although the 
R\'enyi entropy calculated by using $p(e_j)$'s
is for sure bigger than the R\'enyi entanglement entropy, it should not be too far from it. For example, 
one can guess that the R\'enyi entropy in this case follows the area law for gapped systems and has 
logarithmic behavior for critical systems. In the rest of this paper, we will show that indeed this argument is correct and the R\'enyi
entropy of the subsystem energy follows closely the R\'enyi entanglement entropy.

%%%%%%%%%%%%%%%%%%%
\section{Subsystem energy probabilities in the XY chain}
%\label{sec2}
%%%%%%%%%%%%%%%%%%%

The Hamiltonian of the XY-chain is  defined as
\begin{eqnarray}\label{HXY1}
\textbf{H}_{XY}=\hspace{7.5cm}\nonumber\\-J\sum_{j=1}^{L}\Big{[}(\frac{1+\gamma}{4})\sigma_{j}^{x}\sigma_{j+1}^{x}+(\frac{1-\gamma}{4})
\sigma_{j}^{y}\sigma_{j+1}^{y}\Big{]} -\frac{h}{2}\sum_{j=1}^{L}\sigma_{j}^{z},\hspace{0.6cm}
\end{eqnarray}
where the $\sigma_{j}^{\alpha}$ ($\alpha=x,y,z$) are Pauli matrices. $J>0$ is the spin coupling, $\gamma$ is 
the anisotropic parameter and, $h$ is the external magnetic field. For $\gamma=1$, 
and $\gamma=0$, the XY model reduces to the Ising spin chain  and XX chain, respectively. The phase diagram of
the model is rich, 
%and depicted in Fig.1. For $\gamma,h\geq0$, 
there are two different critical lines 
with different universality classes corresponding to the central charges of $c=1$ and $c=\frac{1}{2}$ for the critical XX line $0\leq h<1$ and the critical XY line $h=1$, respectively. 
%%%%%%%%%%%%%%%%%%%%%%%%%%%%%%%%%%%%%%%%%%%%%%%%%
%\begin{figure} [t] 
%\centering
%\includegraphics[width=0.5\textwidth,angle =0]{phase_area_law.eps}
%\caption{(Color online) The phase diagram of the transverse field  XY chain.}  
%\label{fig:XYpahse space}
%\end{figure}

To calculate the statistics of the subsystem energy we first need to write the above Hamiltonian in a more suitable form.
Introducing canonical spinless fermions through the Jordan-Wigner transformation,  $c_l^{\dagger}=\prod_{n<l}\sigma_n^z\sigma_l^{+}$, 
the Hamiltonian \eqref{HXY1} becomes 
\begin{eqnarray}\label{H1}\
\textbf{H}=\textbf{c}^{\dagger}.\hat{\textbf{A}}.\textbf{c}+\frac{1}{2}\textbf{c}^{\dagger}.\hat{\textbf{B}}.\textbf{c}^{\dagger}+\frac{1}{2}\textbf{c}.\hat{\textbf{B}}^{T}.\textbf{c}-\frac{1}{2}{\rm Tr}{\hat{\textbf{A}}},
\end{eqnarray}
with appropriate $\hat{\textbf{A}}$ and $\hat{\textbf{B}}$ matrices. We use the hat symbol to indicate the matrices for the total system, while the symbols 
without hat will indicate the subsystem. The method that we present 
here is quite general and can be used for any Hamiltonian with $\hat{\textbf{A}}$ and $\hat{\textbf{B}}$ being 
symmetric and anti-symmetric matrices\cite{footnote3}.
For the truncated Hamiltonian $\textbf{H}_{D}$ (for quantum chains $D=l$), the same form of the Hamiltonian can be used.
Although the method can be generalized for more general states, we consider that the total system is in its ground state.
To this end, the probability of the subsystem in different energy states can be calculated from the reduced density matrix 
by using the standard method of Refs. \cite{Peschel,Barthel2006,Peschel2009}. The starting point is 
the following reduced density matrix \cite{Barthel2006} written in fermionic coherent basis: 
\begin{eqnarray}\label{rhod1}
\rho_D(\boldsymbol{\xi},\boldsymbol{\xi'})&=&<\boldsymbol{\xi}|\rho_D|\boldsymbol{\xi'}>\nonumber\\
&=&\det\frac{1}{2}(\mathbb{I}-\textbf{G})e^{\frac{1}{2}(\bar{\boldsymbol{\xi}}-\boldsymbol{\xi}')^T
\textbf{F}(\bar{\boldsymbol{\xi}}+\boldsymbol{\xi}')},
\end{eqnarray}
where  we have $\textbf{F}=(\textbf{G}+\mathbb{I})(\mathbb{I}-\textbf{G})^{-1}$ with $\textbf{G}$ being the correlation matrix with the elements $G_{ij}=\langle (c_{i}^{\dagger}-c_{i})(c_{j}^{\dagger}+c_{j})\rangle$. In the above, we introduced the fermionic coherent state as follows,
\begin{eqnarray}\label{fermionic coherent states1}
 |\boldsymbol{\xi}\rangle= |\xi_1,\xi_2,...,\xi_{|D|}\rangle= e^{-\sum_{k=1}^{|D|}\xi_{_{k}}c_k^{\dagger}}|0\rangle,
\end{eqnarray}
where $\xi_k$'s are Grassmann numbers which satisfy the following properties: $\xi_n\xi_m+\xi_m\xi_n=0$ and $\xi_n^2=\xi_m^2=0$. Then, 
it is straightforward to show that
\begin{eqnarray}\label{fermionic coherent states2}
c_{k}|\boldsymbol{\xi}>= \xi_k|\boldsymbol{\xi}>.
\end{eqnarray}
 If we expand the exponential (\ref{rhod1}), we can rewrite it as,
\begin{eqnarray}\label{rhod2}
\rho_D(\boldsymbol{\xi},\boldsymbol{\xi'})=
\det\frac{1}{2}(\mathbb{I}-\textbf{G})e^{\frac{1}{2}\bar{\boldsymbol{\xi}}
\textbf{F}\bar{\boldsymbol{\xi}}}\,\,e^{\frac{1}{2}
\bar{\boldsymbol{\xi}}(\textbf{F}+\textbf{F}^T)\boldsymbol{\xi}'}\,\,e^{-\frac{1}{2}\boldsymbol{\xi}'\textbf{F}\boldsymbol{\xi}'},\hspace{1cm}
\end{eqnarray}
We notice that, $\frac{\bar{\boldsymbol{\xi}}\textbf{F}\bar{\boldsymbol{\xi}}-\bar{\boldsymbol{\xi}}\textbf{F}^T\bar{\boldsymbol{\xi}}}{2}=\bar{\boldsymbol{\xi}}\textbf{F}\bar{\boldsymbol{\xi}}$. 
Then, we get,
\begin{eqnarray}\label{rhod3}
\rho_D(\boldsymbol{\xi},\boldsymbol{\xi'})=
\det\frac{1}{2}(\mathbb{I}-\textbf{G})e^{\frac{1}{2}\bar{\boldsymbol{\xi}}\textbf{F}_{a}\bar{\boldsymbol{\xi}}}\,
\,e^{\frac{1}{2}\bar{\boldsymbol{\xi}}\textbf{F}_{s}\boldsymbol{\xi}'}\,\,e^{-\frac{1}{2}\boldsymbol{\xi}'\textbf{F}_{a}\boldsymbol{\xi}'},\hspace{1cm}
\end{eqnarray}
where $\textbf{F}_{a}=\frac{\textbf{F}-\textbf{F}^T}{2}$ and $\textbf{F}_{s}=\frac{\textbf{F}+\textbf{F}^T}{2}$. By converting 
the Grassmann variables to fermionic operators, we have,
\begin{eqnarray}\label{rhod4}
\rho_D=\det\frac{1}{2}(\mathbb{I}-\textbf{G})e^{\frac{1}{2}\textbf{c}^{\dagger}\textbf{F}_{a}\textbf{c}^{\dagger}}\,\,e^{\frac{1}{2}\textbf{c}^{\dagger}\ln(\textbf{F}_{s}) \textbf{c}}\,\,e^{-\frac{1}{2}\textbf{c}\textbf{F}_{a}\textbf{c}},
\end{eqnarray}
We would like to write the reduced density matrix as
\begin{eqnarray}\label{entanglement Hamiltonian}
\rho_{D}=\det\frac{1}{2}(\mathbb{I}-\textbf{G})e^{\mathcal{H}}
\end{eqnarray}
Where $\mathcal{H}$ is the entanglement Hamiltonian that can be calculated by combining 
the exponential terms in the equation ~\ref{rhod4}. In other words, we would like to have
\begin{widetext}
\begin{eqnarray}\label{H2}\
\mathcal{H}=\sum_{lm}[c_{l}^{\dagger}M_{lm}c_{m}+\frac{1}{2}c_{l}^{\dagger}N_{lm}c_{m}^{\dagger}+\frac{1}{2}c_{l}N_{ml}c_{m}]-\frac{1}{2}{\rm Tr}{\textbf{M}},
\end{eqnarray} 
\end{widetext}
where $\textbf{M}$ and $\textbf{N}$ are symmetric and antisymmetric matrices. They can be calculated using Balian-Brezin formula \cite{Balian1969}
\begin{widetext}
\begin{eqnarray}\label{mat_T}\
\textbf{T}=e^{\begin{pmatrix}
\textbf{M} & \textbf{N}\\
\textbf{-N} & \textbf{-M}\\
  \end{pmatrix}}
  = \begin{pmatrix}
\textbf{T}_{11} & \textbf{T}_{12}\\
\textbf{T}_{21} & \textbf{T}_{22}\\
  \end{pmatrix}=
 \begin{pmatrix}
\textbf{F}_{s}-\textbf{F}_{a}\textbf{F}_{s}^{-1}\textbf{F}_{a} & \textbf{F}_{a}\textbf{F}_{s}^{-1}\\
-\textbf{F}_{s}^{-1}\textbf{F}_{a}& \textbf{F}_{s}^{-1}\\
  \end{pmatrix}. 
\end{eqnarray} 
\end{widetext}
Then, we get
\begin{eqnarray}\label{mat_T3}\
\begin{pmatrix}
\textbf{M} & \textbf{N}\\
\textbf{-N} & \textbf{-M}\\
  \end{pmatrix}=
\ln \begin{pmatrix}
\textbf{F}_{s}-\textbf{F}_{a}\textbf{F}_{s}^{-1}\textbf{F}_{a} & \textbf{F}_{a}\textbf{F}_{s}^{-1}\\
-\textbf{F}_{s}^{-1}\textbf{F}_{a}& \textbf{F}_{s}^{-1}\\
  \end{pmatrix}.
\end{eqnarray} 
Now, we can rewrite the entanglement Hamiltonian as follows
\begin{eqnarray}\label{H2}\
\mathcal{H}=\frac{1}{2}(\textbf{c}^{\dagger}\,\,\textbf{c})\begin{pmatrix}
\textbf{M} & \textbf{N}\\
-\textbf{N} & -\textbf{M}\\
  \end{pmatrix}
  \begin{pmatrix}
\textbf{c}\\
\textbf{c}^{\dagger}\\
  \end{pmatrix}+\frac{1}{2}{\rm Tr}\ln{(\textbf{F}_{s})}.
\end{eqnarray} 
The second step is to diagonalize the truncated  Hamiltonian $\textbf{H}_{D}$ with the standard method of Ref.\cite{LSM}, see Appendix A. 
The idea is based on canonical transformation 
\begin{eqnarray}\label{mat_U1}\
\begin{pmatrix}
\textbf{c} \\
\textbf{c}^{\dagger} \\
\end{pmatrix}=
\textbf{U}^{\dagger}
\begin{pmatrix}
\boldsymbol{\eta} \\
\boldsymbol{\eta}^{\dagger} \\
\end{pmatrix}.
\end{eqnarray}
By using the above relation , we can rewrite the entanglement Hamiltonian with respect to the $\eta_{k}$'s as follows:
\begin{eqnarray}\label{H3}\
\mathcal{H}&=&\frac{1}{2}(\boldsymbol{\eta}^{\dagger}\,\,\boldsymbol{\eta}) \textbf{U}
  \begin{pmatrix}
\textbf{M} & \textbf{N}\\
-\textbf{N} & -\textbf{M}\\
  \end{pmatrix}\textbf{U}^{\dagger}
  \begin{pmatrix}
\boldsymbol{\eta}\\
\boldsymbol{\eta}^{\dagger}\\
  \end{pmatrix}+\frac{1}{2}{\rm Tr}\ln{(\textbf{F}_{s})}\nonumber\\
  &=&
\frac{1}{2}(\boldsymbol{\eta}^{\dagger}\,\,\boldsymbol{\eta})\textbf{Q}
  \begin{pmatrix}
\boldsymbol{\eta}\\
\boldsymbol{\eta}^{\dagger}\\
  \end{pmatrix}+\frac{1}{2}{\rm Tr}\ln{(\textbf{F}_{s})},
\end{eqnarray} 
where
\begin{eqnarray}\label{H3}\
\textbf{Q}=\begin{pmatrix}
\textbf{Q}_{11} & \textbf{Q}_{12}\\
\textbf{Q}_{21} & \textbf{Q}_{22}\\
  \end{pmatrix}= \textbf{U}
  \begin{pmatrix}
\textbf{M} & \textbf{N}\\
-\textbf{N} & -\textbf{M}\\
  \end{pmatrix}\textbf{U}^{\dagger}.
\end{eqnarray} 
The reduced density matrix in the $\eta$ basis finally becomes:
\begin{eqnarray}\label{rhod5}
\rho_{D}=
\det\frac{1}{2}(\mathbb{I}-\textbf{G})[{\rm \det}(\textbf{F}_{s})]^{\frac{1}{2}}e^{\frac{1}{2}(\boldsymbol{\eta}\,\,\boldsymbol{\eta}^{\dagger})\begin{pmatrix}
\textbf{Q}_{11} & \textbf{Q}_{12}\\
\textbf{Q}_{21} & \textbf{Q}_{22}\\
  \end{pmatrix}
  \begin{pmatrix}
  \boldsymbol{\eta}\\
\boldsymbol{\eta}^{\dagger}\\
  \end{pmatrix}}.\hspace{0.5cm}
\end{eqnarray}
Now, we define the coherent basis of the $\eta$ representation as
\begin{eqnarray}\label{fermionic coherent states2}
\eta_{k}|\boldsymbol{\gamma}>= \gamma_k|\boldsymbol{\gamma}>.
\end{eqnarray}
Using the above basis, we have
\begin{widetext}
\begin{eqnarray}\label{rhod6}
\langle \boldsymbol{\gamma} |\rho_{D}|\boldsymbol{\gamma}'\rangle=\det\frac{1}{2}(\mathbb{I}-\textbf{G})[{\rm \det}(\textbf{F}_{s})]^{\frac{1}{2}}\langle \boldsymbol{\gamma} | e^{\frac{1}{2}(\boldsymbol{\eta}^{\dagger}\,\,\boldsymbol{\eta})\begin{pmatrix}
\textbf{Q}_{11} & \textbf{Q}_{12}\\
\textbf{Q}_{21} & \textbf{Q}_{22}\\
  \end{pmatrix}
  \begin{pmatrix}
\boldsymbol{\eta}\\
\boldsymbol{\eta}^{\dagger}\\
  \end{pmatrix}}|\boldsymbol{\gamma}'\rangle.
\end{eqnarray}
\end{widetext}
To calculate the above equation we first define $\tilde{\textbf{T}}$ matrix,
\begin{eqnarray}\label{Ttildeha}
\tilde{\textbf{T}}=e^{\textbf{Q}}=\textbf{U}\begin{pmatrix}
\textbf{F}_{s}-\textbf{F}_{a}\textbf{F}_{s}^{-1}\textbf{F}_{a} & \textbf{F}_{a}\textbf{F}_{s}^{-1}\\
-\textbf{F}_{s}^{-1}\textbf{F}_{a}& \textbf{F}_{s}^{-1}\\
  \end{pmatrix}\textbf{U}^{\dagger}.
\end{eqnarray}
and
\begin{eqnarray}\label{matrixes}\
\tilde{\textbf{X}}&=&\tilde{\textbf{T}}_{12}(\tilde{\textbf{T}}_{22})^{-1},\nonumber\\
\tilde{\textbf{Z}}&=&(\tilde{\textbf{T}}_{22}^{-1})\tilde{\textbf{T}}_{21},\nonumber\\
e^{-\tilde{\textbf{Y}}}&=&\tilde{\textbf{T}}_{22}^{T}.
\end{eqnarray} 
Then, by decomposing the exponential factor in the equation \ref{rhod6} (using the Balian-Brezin formula \cite{Balian1969}), we get
\begin{eqnarray}\label{rhod7}
\langle \boldsymbol{\gamma} |\rho_{D}|\boldsymbol{\gamma}'\rangle
=\det\frac{1}{2}(\mathbb{I}-\textbf{G})[{\rm \det}(\textbf{F}_{s})]^{\frac{1}{2}}\hspace{2cm}\nonumber\\
\times\langle \boldsymbol{\gamma} | e^{\frac{1}{2}\boldsymbol{\eta}^{\dagger}\tilde{\textbf{X}}\boldsymbol{\eta}^{\dagger}}\,\,e^{\boldsymbol{\eta}^{\dagger}\tilde{\textbf{Y}}\boldsymbol{\eta}}\,\,e^{\frac{1}{2}\boldsymbol{\eta}\tilde{\textbf{Z}}\boldsymbol{\eta}}\,\,e^{-\frac{1}{2}{\rm Tr}{\tilde{\textbf{Y}}}}|\boldsymbol{\gamma}'\rangle,
\end{eqnarray}
which becomes,
\begin{eqnarray}\label{rhod8s1}
\langle \boldsymbol{\gamma} |\rho_{D}|\boldsymbol{\gamma}'\rangle=
\det\frac{1}{2}(\mathbb{I}-\textbf{G})\hspace{4cm}\nonumber\\
\times[{\rm \det}(\textbf{F}_{s})]^{\frac{1}{2}}\,\,
 e^{\frac{1}{2}\bar{\boldsymbol{\gamma}}\tilde{\textbf{X}}\bar{\boldsymbol{\gamma}}}\,\,
 e^{\bar{\boldsymbol{\gamma}}e^{\tilde{\textbf{Y}}}\boldsymbol{\gamma'}}\,\,
 e^{\frac{1}{2}\boldsymbol{\gamma'}\tilde{\textbf{Z}}\boldsymbol{\gamma'}}\,\,
 e^{-\frac{1}{2}{\rm Tr}{\tilde{\textbf{Y}}}}.\hspace{1cm}
\end{eqnarray}
After defining
\begin{eqnarray}\label{BB_rel3}
\tilde{\textbf{F}}_{s}&=&e^{\tilde{\textbf{Y}}},\nonumber\\
\tilde{\textbf{F}}&=&\tilde{\textbf{F}}_{s}+\tilde{\textbf{F}}_{a}=\tilde{\textbf{X}}+e^{\tilde{\textbf{Y}}},
\end{eqnarray}
 the reduced density matrix becomes,
 \begin{widetext}
\begin{eqnarray}\label{rhod8s2}
\langle \boldsymbol{\gamma} |\rho_{D}|\boldsymbol{\gamma}'\rangle=
\det\frac{1}{2}(\mathbb{I}-\textbf{G})[{\rm \det}(\textbf{F}_{s})]^{\frac{1}{2}} e^{-\frac{1}{2}{\rm Tr}\tilde{\textbf{Y}}}\,\,e^{\frac{1}{2}(\bar{\boldsymbol{\gamma}}-\boldsymbol{\gamma'})
\tilde{\textbf{F}}(\bar{\boldsymbol{\gamma}}+\boldsymbol{\gamma'})}
=\det\frac{1}{2}(\mathbb{I}-\textbf{G})[\frac{{\rm \det}(\textbf{F}_{s})}{{\rm \det}(\tilde{\textbf{F}}_{s})}]^{\frac{1}{2}} \,\,e^{\frac{1}{2}(\bar{\boldsymbol{\gamma}}-\boldsymbol{\gamma'})\tilde{\textbf{F}}(\bar{\boldsymbol{\gamma}}+\boldsymbol{\gamma'})}.
\end{eqnarray}
\end{widetext}
At this point, we explain how one can use the above equation to calculate the desired probabilities. The procedure is similar to the calculation of formation probabilities\cite{Najafi2016}.
First of all, one can think  of $|\boldsymbol{\gamma}\rangle$ as a coherent state corresponding to the different excitation modes. For example,
when all the $\gamma_k$'s are zero the corresponding coherent state is equal to the vacuum (no excited modes). In this case, it is easy to see that to find the probability of a subsystem in its ground state, one needs to put all the $\gamma$'s equal to zero, then
\begin{eqnarray}\label{Ttilde}
p(e_g)=\det\frac{1}{2}(\mathbb{I}-\textbf{G})[\frac{{\rm \det}(\textbf{F}_{s})}{{\rm \det}(\tilde{\textbf{F}}_{s})}]^{\frac{1}{2}}.
\end{eqnarray}
To find the  probability of other energies, one needs to know the corresponding modes $\lambda_k$'s in 
which generate the desired energy. This means in the corresponding coherent state we put one fermion in the $\eta$ basis which simply is equivalent to the following Grassmann integration
\begin{eqnarray}\label{grassmann integration}
|0,0,...,1_k,0,...,0\rangle=\int d\gamma_k |\boldsymbol{\gamma}\rangle.
\end{eqnarray}
The left hand side is the excited state with the mode $k$ excited. It should now be clear that if we want to calculate the probability of 
the corresponding state we just need to first put $\boldsymbol{\gamma}'=\boldsymbol{\gamma}$ and then put all the 
modes that are not excited equal to zero and then Grassmann integrate over $\gamma_k$. This will lead to the one of the principal minors of the matrix $\tilde{\textbf{F}}$.
The procedure is the same when we excite more modes that lead to the other excited states. There are $2^{|D|}$ possible mode excitations which correspond
to the same number of principal minors that the matrix $\tilde{\textbf{F}}$ have. One should note that sometimes different mode excitations lead to
the same energy for the excited state. In this cases, we need to sum the  minors. One can now summarize the main equation of the article as
\begin{eqnarray}\label{main}
p(e)=\det\frac{1}{2}(\mathbb{I}-\textbf{G})[\frac{{\rm \det}(\textbf{F}_{s})}{{\rm \det}(\tilde{\textbf{F}}_{s})}]^{\frac{1}{2}}\sum_e \text{Min} [\tilde{\textbf{F}}],
\end{eqnarray}
where the sum is over the degeneracy of the energy $e$. In the end, in Appendix B, we 
drive an explicit formula for the generating function of the subsystem energy.

%%%%%%%%%%%%%%%%%%%
\section{ R\'enyi entropy of the subsystem energy}
%\label{sec2}
%%%%%%%%%%%%%%%%%%%

\begin{figure} [t] 
\includegraphics[width=0.45\textwidth,angle =0]{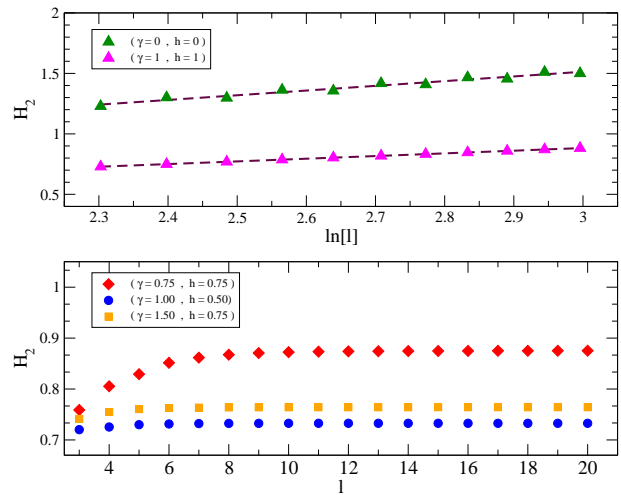}
\caption{(color online)Up: logarithmic behavior of the R\'enyi entropy, $\alpha=2$, of the subsystem energy of the critical XY chain for $\gamma=h=0$ and $\gamma=h=1$. 
Down: Area law of the R\'enyi entropy for various values of the parameters of the gapped XY chain. } 
\label{Renyi2}
\end{figure}

Using the equation (\ref{main}), we first, calculated the R\'enyi entropy of the subsystem energy for different gapped points of an infinite XY chain and verified 
the area law for $\alpha\geq \alpha_c$, see Figure 1; where $\alpha_c$ seems to be close to one\cite{footnotealphac}.
Then, we calculated the same quantity for the critical regions which shows the logarithmic behavior for the R\'enyi entropy, see Figure \ref{Renyi2}. 
%%%%%%%%%%%%%%%%%%%%%%%%%%%%%%%%%%%%%%%%%%%%%%%%%
%%%%%%%%%%%%%%%%%%%%%%%%%%%%%%%%%%%%%%%%%%%%%%%%%%%%%%%%%%%%%%%%%%%%%%

\begin{figure} [t] 
\includegraphics[width=0.45\textwidth,angle =0]{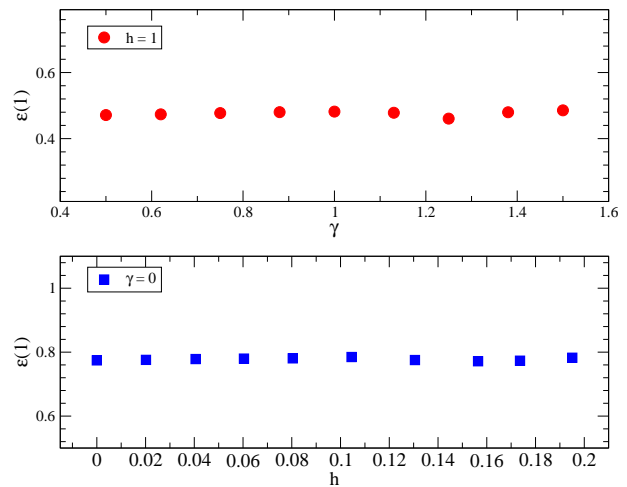}
\caption{(color online) Universality of the coefficient of the logarithm, $\epsilon(1)$ on the critical XY line, i.e. $h=1$ and critical XX line, i.e. $\gamma=0$. Here, the fits are performed using the subsystem sizes  $8\leq l\leq16$. } 
\label{universal}
\end{figure}

\begin{figure} [t] 
\includegraphics[width=0.45\textwidth,angle =0]{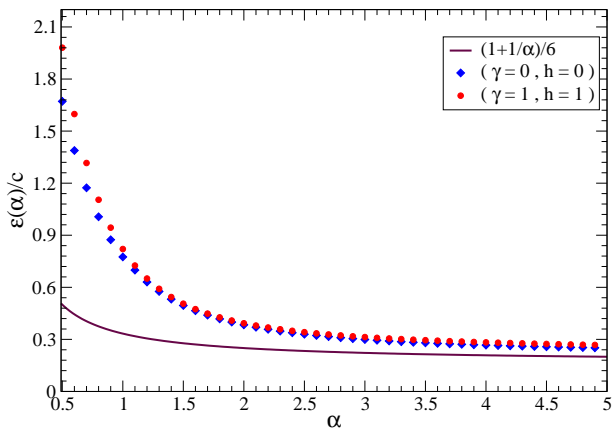}
\caption{(color online)$\frac{\epsilon(\alpha)}{c}$ for the critical XX, i.e. $\gamma=h=0$ and the critical Ising, i.e. $\gamma=h=1$ with respect to $\alpha$. The solid line is the CFT result for the R\'enyi entanglement entropy. Here, the fits are performed using the subsystem sizes $8\leq l\leq16$.} 
\label{epsilonalpha}
\end{figure}

In general, for the R\'enyi entropy, we find
\begin{equation}\label{Renyi critical}
H_{\alpha}=\epsilon(\alpha)\ln l+\beta_{\alpha},
\end{equation}
for $l\geq 6$. Afterwards, in the Figure \ref{universal}, we verified that these coefficients are universal in the sense that on the critical XX and XY lines, their values do not change significantly. We note that there is
a freedom in choosing the boundary conditions of the  truncated  Hamiltonian. To check that these numbers are insensitive to these boundary conditions, we also considered periodic boundary conditions
for $H_l$ and repeated the calculations and reached to the same numbers. 
Note that in calculating the Shannon (R\'enyi) entropy, we use 
all the probabilities including those that are associated with very high energies in the subsystem. These states which are far from the ground state normally do not show universal
behavior and one can not describe them by quantum field theories. Our calculations indicates that, in the scaling limit, most probably the contribution of these states in the calculation of the Shannon  entropy is negligible\cite{footnotesmallP}.  In the Figure \ref{epsilonalpha}, we plotted the  $\frac{\epsilon(\alpha)}{c}$ and the coefficient of the logarithm in the equation (\ref{Renyi conformal}) for the XX and critical Ising chain. Remarkably, they follow a very similar behavior which confirms our original discussion regarding the closeness of the local energy basis to the Schmidt basis. 
%%%%%%%%%%%%%%%%%%%%%%%%%%%%%%%%%%%%%%%%%%%%%%%%%
%%%%%%%%%%%%%%%%%%%%%%%%%%%%%%%%%%%%%%%%%%%%%%%%%%%%%%%%%%%%%%%%%%%%%%

It also indicates that probably (at least for large $\alpha$'s) the coefficient $\epsilon(\alpha)$ is linearly proportional to the central charge. For small $\alpha$'s the deviations between the three graphs are more significant which should not be surprising since here, the more important probabilities are the smaller ones which are related to the very high excited states of the subsystem Hamiltonian. We also noticed that in the regime $\alpha<0.5$ for the considered sizes $l\leq 20$ we do not see a nice logarithmic behavior.
%%%%%%%%%%%%%%%%%%%%%%%%%%%%%%%%%%%%%%%%%%%%%%%%%%%%%
%%%%%%%%%%%%%%%%%%%%%%%%%%%%%%%%%%%%%%%%%%%%%%%%%
Since the number of possible energies for the subsystem increases exponentially with the subsystem size there is a limitation in calculating the 
Shannon entropy for large subsystems.  That makes the estimation of the coefficient $\epsilon(\alpha)$ very difficult. 
Instead of the Shannon entropy if one considers $\alpha\to\infty$, then, the only probability that one needs to take into account is the probability of the subsystem being in the 
ground state which is the largest probability. In this case, one can go to relatively large sizes about $l=1000$ and check all the conclusions with much more accuracy. 
Indeed, we were able to calculate the coefficient of the logarithm in this case with high accuracy for the XX and critical Ising chain:
 %%%%
\begin{equation}\label{infinite Renyi}
\epsilon_{XX}(\infty)=0.222(2),\hspace{1cm}\epsilon_{Ising}(\infty)=0.112(2).
\end{equation}
%%%%%%%
We also checked the universality of these values on the critical line, see Appendix C. For the semi-infinite systems, the coefficients are  within one percent 
from the half of the above values. Note that this quantity is different from the quantity called fidelity in Refs.\cite{Dubail2011,Dubail2013}. In the following we further support our ideas by first introducing the truncated Shannon and compare it with the entanglement entropy of the subsystem energy. Furthermore, we provide comprehensive numerical results on the closeness of  R\'enyi entanglement entropy and R\'enyi entropy of the subsystem energy and Schmidt basis and the truncated Hamiltonian basis.

%%%%%%%%%%%%%%%%%%%%%%%%%%%%%%%%%%%%%%%%%%%%%%%%%%%%%%%%%%%%%%%%%%%%%%%%%%%%%%%%%%%%%%%%
\subsection{ Truncated Shannon and entanglement entropy of the subsystem energy }
\begin{figure} [t] 
\includegraphics[width=0.45\textwidth,angle =0]{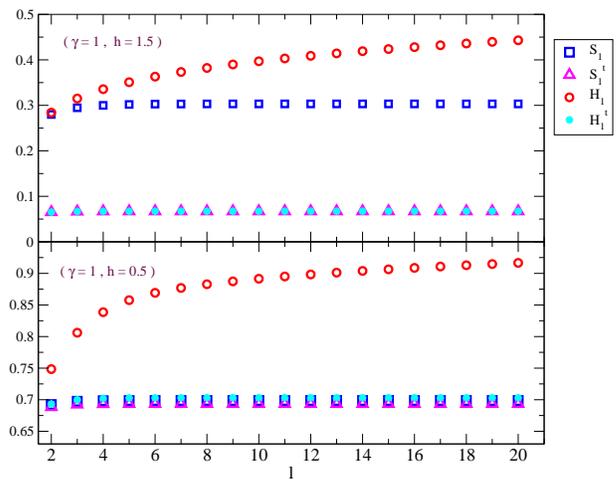}
\caption{(color online)Comparision of the (truncated) Shannon entropy $(H^t_1)H_1$
with the (truncated) von Neumann entropy $(S^t_1)S_1$ for two gapped points $(\gamma,h)=(1,\frac{1}{2})$ and $(\gamma,h)=(1,\frac{3}{2})$.}  
\label{shan_trun}
\end{figure}

%%%%%%%%%%%%%%%%%%%%%%%%%%%%%%%%%%%%%%%%%%%%%%%%%%%%%%%%%%%%%%%%%%%%%%%%%%%%%%%%%%%%%%%%
All of the results that we presented so far indicate the closeness of our probabilities $p_j$ to the Schmidt coefficients $\lambda_j$. Most interestingly similar to the Schmidt coefficients 
we realized that just the first few probabilities are big and the rest are very small\cite{footnotesmall}. To show how much these probabilities are close 
to the Schmidt coefficients and see their contribution to the full Shannon entropy it is useful to define a truncated 
Shannon and truncated von Neumann entropy \cite{footnotetruncatedRenyi} as
$H_{1}^t=-\sum_{j=1}^kp_j\ln p_j$ and $S_{1}^t=-\sum_{j=1}^k\lambda_j\ln \lambda_j$, where for our model $k=1$($2$) for $h>1$($h<1$). We note that here
the change in the number $k$ is consistent with the paramagnetic-ferromagnetic phase transition.
Figure \ref{shan_trun} shows that although $H_1$ is not very close to the $S_1$, the agreemet between $H^t_1$ and $S^t_1$ is striking\cite{footnoteothermodels}. 
This calculation means that
although compare to the first important probabilities the other probabilities 
are exponentially small, their decay is not as strong as the decay in the Schmidt coefficients. We will do further investigation on the next subsection. 
%Finally, it is worth mentioning that for finite systems the Shannon information has one %important difference from the von Neumann entropy. Because of the 
%Schmidt decomposition for the von Neumann entanglement entropy we have $S(l)=S(L-l)$, %however, this relation is not valid (weakly broken) for the Shannon entropy  that we %considered here. 
%We will discuss in more detail the finite systems in another work. 

%%%%%%%%%%%%%%%%%%%%%%%%%%%%%%%%%%%%%%%%%%%%%%%%%
%%%%%%%%%%%%%%%%%%%%%%%%%%%%%%%%%%%%%%%%%%%%%%%%%%%%%%%%%%%%%%%%%%%%%%

%%%%%%%%%%%%%%%%%%%%%%%%%%%%%%%%%%%%%%%%%%%%%%%%%%%%%%%%%%%%%%%%%%%%%%%%%%%%%%%%%%%%%%%%
\subsection{ R\'enyi entanglement entropy and R\'enyi entropy of the subsystem energy }
%%%%%%%%%%%%%%%%%%%%%%%%%%%%%%%%%%%%%%%%%%%%%%%%%%%%%%%%%%%%%%%%%%%%%%%%%%%%%%%%%%%%%%%%

%%%%%%%%%%%%%%%%%%%%%%%%%%%%%%%%%%%%%%%%%%%%%%%%5%%%%%%%%%%%%%%%%%%%%%%%%%
\begin{figure}[hthp!]
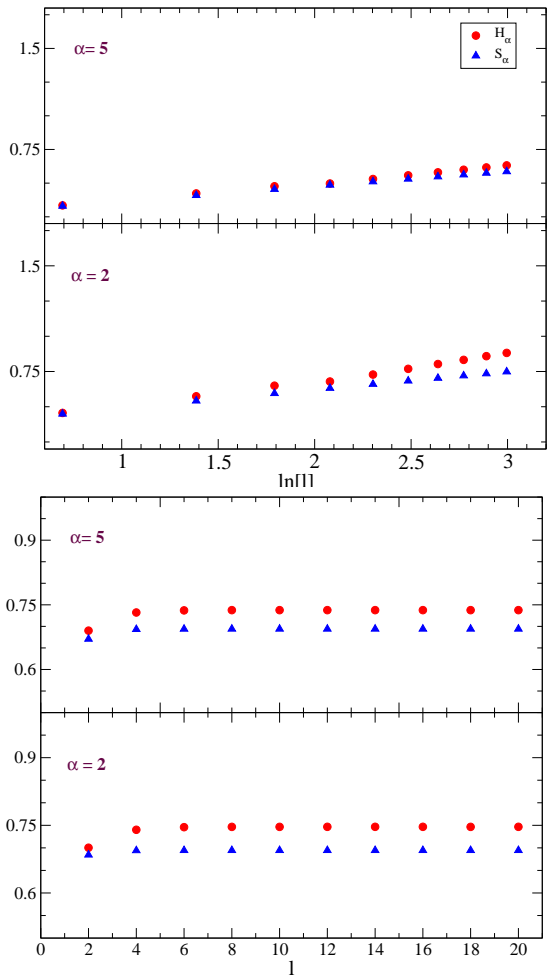

  \centering
    \includegraphics[width=0.40\textwidth]{SE_a1h1.eps}\hspace{1cm}
    \includegraphics[width=0.40\textwidth]{SE_a1h05.eps}
    \label{EntShan}
\caption{(color online) Comparing the entanglement entropy and  the R\'enyi entropy of the subsystem energy for  $\alpha=2$ and $5$. 
Top: Critical XY chain $(\gamma=1, h=1)$. In this case,  to better verify the closeness, we have depicted the entropies in the semi-log plots which indicates that for bigger 
$\alpha$'s, they become closer. Bottom: Non-critical XY chain $(\gamma=1,h=0.5)$.} 
\label{fig_SE}
\end{figure}
%%%%%%%%%%%%%%%%%%%%%%%%%%%%%%%%%%%%%%%%%%%%%%%%%%%%%%%%%%%%%%%%%%%

\begin{figure} [t] 
\includegraphics[width=0.5\textwidth,angle =0]{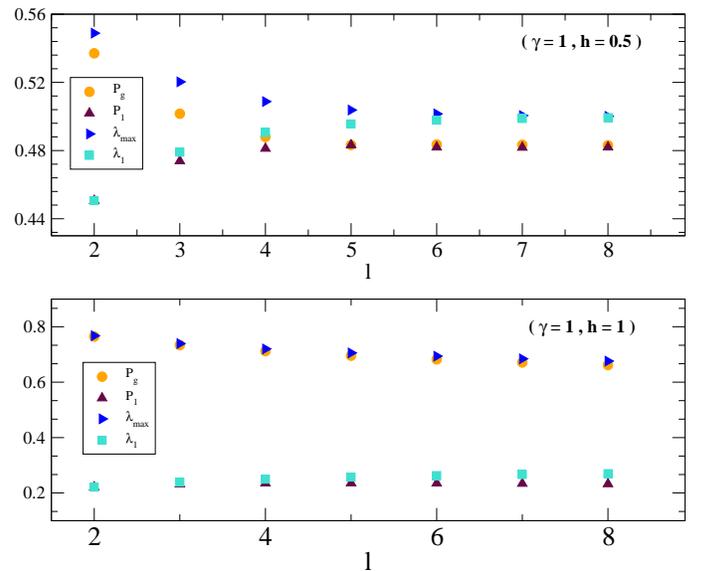}
\centering
\caption{(color online) Comparing the two biggest probabilities in the subsystem energy basis indicated as $P_{g}$ and $P_{1}$ with the two biggest eigenvalues of the reduced density matrix, indicated with $\lambda_{max}$ and $\lambda_1$. Both for the critical $(\gamma=1,h=1)$ and the noncritical case $(\gamma=1,h=0.5)$ depicted in the bottom and the top respectively. The $P_{g}$ closely mimics $\lambda_{max}$ while $P_{1}$ mimics the $\lambda_{1}$.} 
\label{fig_Plambda}
\end{figure}

In this subsection, we provide some numerical results to support our claims on the closeness of the R\'enyi entanglement 
entropy and the R\'enyi entropy of the subsystem energy. There are at least a  couple of different ways to investigate 
this matter. The first and more direct one is to compare the two entropies for different $\alpha$'s. In the Figure \ref{fig_SE}, 
we have depicted the two entropies, $H_{\alpha}$ and $S_{\alpha}$ with respect to $l$ (the size of the subsystem for the critical Ising model) 
for various values of $\alpha=2$ and $5$. It is clear that for bigger $\alpha$'s, the two entropies mimic each other closely. Similar 
conclusions are also valid for the other critical systems. We have done similar calculations for the non-critical ground states 
as well and the result is shown in the Figure \ref{fig_SE}. The conclusions are similar which support the closeness of the R\'enyi 
entropy of the subsystem energy to the R\'enyi entanglement entropy. 
Since for the bigger $\alpha$'s the convergence is better ( even for the small sites considered here), one immediately guesses 
that the closeness of the R\'enyi entropy of the subsystem energy to the R\'enyi entanglement entropy might be due to the closeness 
of the biggest probabilities to the largest eigenvalues of the reduced density matrix. To verify this argument, we directly compare 
the two sets of numbers as the second method to see the similarity between these two quantities. 
In the Figure \ref{fig_Plambda}, we numerically showed that this is actually true. In other words, at least, the two biggest 
probabilities, i.e. $P_g$ and $P_1$, closely follow the two biggest eigenvalues of the reduced density 
matrix, i.e. $\lambda_{\text{max}}$ and $\lambda_1$. Furthermore, we have checked numerically in many 
other points of the phase diagram and observed the same behavior. It might be interesting to investigate 
the connection of these conclusions to the behavior of the Schmidt gap under the phase transition studied recently in \cite{DeChiara2012,Bayat2014}.

%%%%%%%%%%%%%%%%%%%%%%%%%%%%%%%%%%%%%%%%%%%%%%%%%%%%%
%%%%%%%%%%%%%%%%%%%%%%%%%%%%%%%%%%%%%%%%%%%%%%%%%

\begin{figure}[hthp!]
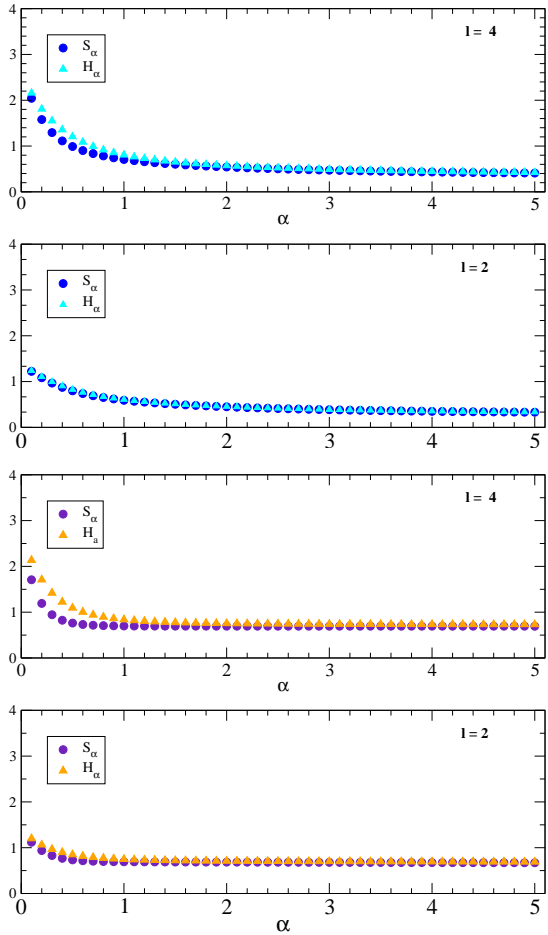

  \centering
    \includegraphics[width=0.4\textwidth]{EntShan_a1h1.eps}\hspace{2cm}
    \includegraphics[width=0.4\textwidth]{EntShan_a1h05.eps}
\caption{(color online) Comparing the two entropies $H_{\alpha}$ and $S_{\alpha}$ with respect to $\alpha$ for small sizes $l=2$ and $l=4$ 
at the critical Ising point $(\gamma=1,h=1)$ depicted in the top and the noncritical point $(\gamma=1,h=0.5)$ depicted in the bottom panel. 
For both cases and for both of the sizes, we observe that the R\'enyi entropy of the subsystem energy behave similarly to the R\'enyi 
entanglement entropy and they get closer for bigger values of $\alpha$. } 
\label{fig_ES}
\end{figure}

Finally, we  have also investigated the closeness of the 
two entropies for small sizes. In the Figure \ref{fig_ES}, we 
illustrate the $H_{\alpha}$ and $S_{\alpha}$ as a function of $\alpha$ for $L=2$ and $L=4$ 
at both the critical and the non-critical points. It is clear that even for small sizes, the R\'enyi entropy 
of the subsystem energy chases the R\'enyi entanglement entropy and becomes incredibly close for bigger values of $\alpha$ as we mentioned earlier. 

%%%%%%%%%%%%%%%%%%%%%%%%%%%%%%%%%%%%%%%%%%%%%%%%5%%%%%%%%%%%%%%%%%%%%%%%%%
%%%%%%%%%%%%%%%%%%%%%%%%%%%%%%%%%%%%%%%%%%%%%%%%%%%%%%%%%%%%%%%%%%

%%%%%%%%%%%%%%%%%%%%%%%%%%%%%%%%%%%%%%%%%%%%%%%%%%%%%%%%%%%%%%
\subsection{Schmidt basis and the truncated Hamiltonian basis}
%%%%%%%%%%%%%%%%%%%%%%%%%%%%%%%%%%%%%%%%%%%%%%%%%%%%%%%%%%%%%%

%%%%%%%%%%%%%%%%%%%%%%%%%%%%%%%%%%%%%%%%%%%%%%%%5%%%%%%%%%%%%%%%%%%%%%%%%%
\begin{figure}[hthp!]
    \includegraphics[width=0.450\textwidth]{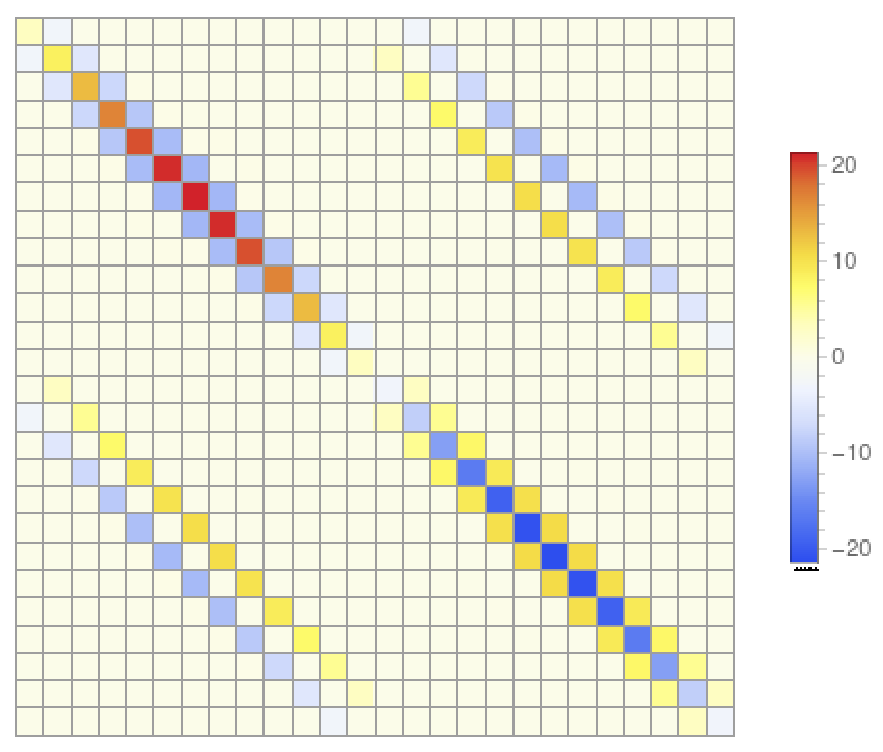}
    \includegraphics[width=0.450\textwidth]{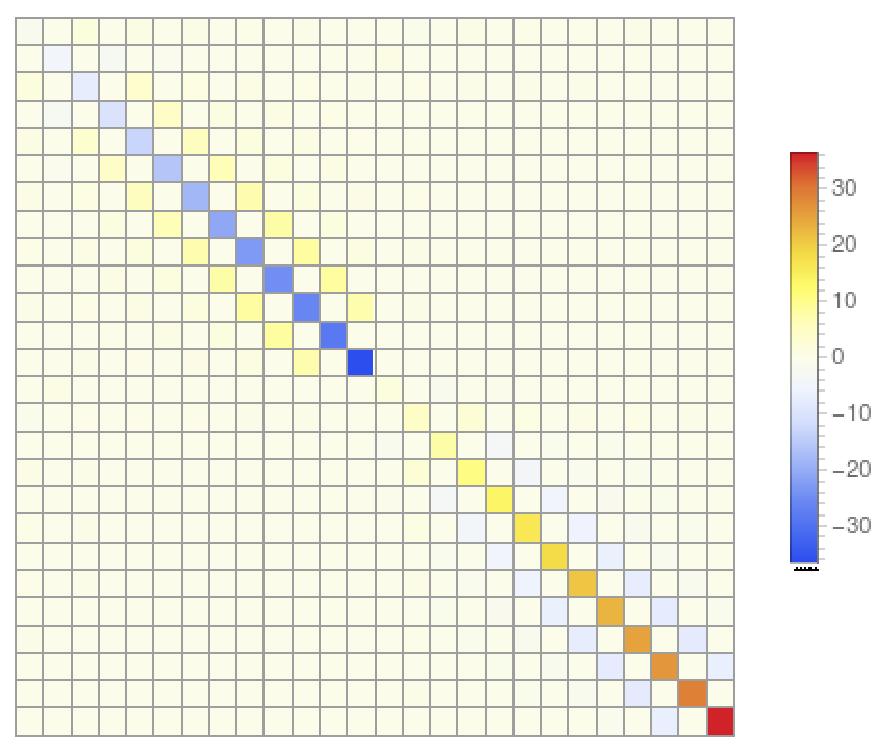}
    \label{EntShan}
\caption{(color online) Comparing matrices $\begin{pmatrix}
\textbf{M} & \textbf{N}\\
-\textbf{N} & -\textbf{M}\\
  \end{pmatrix}$ illustrated in the top and $\textbf{Q}= \textbf{U}\begin{pmatrix}
\textbf{M} & \textbf{N}\\
-\textbf{N} & -\textbf{M}\\
  \end{pmatrix}\textbf{U}^{\dagger}$ illustrated in the bottom for a subsystem with size $l=13$. Notice that both matrices are block matrices of size $2l$ and the majority of the matrix elements are nearly zero. Although there are some nonzero values in the diagonal part of the  matrix ${\textbf{N}}$ which manifest itself in the upper right and lower left diagonal of the matrix in the bottom panel, it is remarkable that after rewriting it in the truncated Hamiltonian basis, those elements nearly vanish and the result matrix $\textbf{Q}$ becomes an almost diagonal matrix. }
\label{fig_NMQ}
\end{figure}
%%%%%%%%%%%%%%%%%%%%%%%%%%%%%%%%%%%%%%%%%%%%%%%%%%%%%%%%%%%%%%%%
Here, we study the difference between the Schmidt basis and the truncated Hamiltonian basis. To this end, we compare the reduced density matrix written in these two bases and show that they are close to each other in the sense of norm distance. The reduced density matrix in the truncated Hamiltonian basis as we discussed in section III, has the following form
\begin{eqnarray}\label{TH basis}
\rho=\det\frac{1}{2}(\mathbb{I}-\textbf{G})[{\rm \det}(\textbf{F}_{s})]^{\frac{1}{2}}e^{\frac{1}{2}(\boldsymbol{\eta}^{\dagger}\,\,\boldsymbol{\eta})\textbf{Q}
  \begin{pmatrix}
  \boldsymbol{\eta}\\
\boldsymbol{\eta}^{\dagger}\\
  \end{pmatrix}}.
\end{eqnarray}
Note that here, we use the word \textit{basis} in the sense of writing the density matrix with respect to the proper creation-annihilation operators. The reason will be clear in few lines.
However, in the Schmidt \textit{basis}, it has the following form
\begin{eqnarray}\label{Schmidt basis}
\rho=\det\frac{1}{2}(\mathbb{I}-\textbf{G})[{\rm \det}(\textbf{F}_{s})]^{\frac{1}{2}}e^{\frac{1}{2}(\boldsymbol{\delta}^{\dagger}\,\,\boldsymbol{\delta})\textbf{D}
  \begin{pmatrix}
  \boldsymbol{\delta}\\
\boldsymbol{\delta}^{\dagger}\\
  \end{pmatrix}},
\end{eqnarray}
where $\textbf{D}$ is a diagonal matrix. The above two density operators written in the current form are the same 
density operators written with respect to the different creation-annihilation operators. Note 
that by diagonalizing the $\textbf{Q}$, we reach to the matrix $\textbf{D}$. Clearly, 
if the $\textbf{Q}$ was diagonal, then, the two bases were equal. However, as a remarkable result, 
the matrix $\textbf{Q}$ becomes close to diagonal. In Figure \ref{fig_NMQ},  we have depicted the $\begin{pmatrix}
\textbf{M} & \textbf{N}\\
-\textbf{N} & -\textbf{M}\\
  \end{pmatrix}$ and $\textbf{Q}= \textbf{U}\begin{pmatrix}
\textbf{M} & \textbf{N}\\
-\textbf{N} & -\textbf{M}\\
  \end{pmatrix}\textbf{U}^{\dagger}$ for the subsystem with the size $l=13$. Comparing these two matrices, it is striking that the $\rho_D$ written in the truncated Hamiltonian becomes near diagonal as the off-diagonal elements washed out in this new basis.
  
 To quantify the difference between the two bases at the level of the density matrices, we first consider that the $\eta$ is equal to $\delta$ in the equation (\ref{TH basis}) and write 
\begin{eqnarray}\label{TH basis 2}
\rho_{e}=\det\frac{1}{2}(\mathbb{I}-\textbf{G})[{\rm \det}(\textbf{F}_{s})]^{\frac{1}{2}}e^{\frac{1}{2}(\boldsymbol{\delta}^{\dagger}\,\,\boldsymbol{\delta})\textbf{Q}
  \begin{pmatrix}
  \boldsymbol{\delta}\\
\boldsymbol{\delta}^{\dagger}\\
  \end{pmatrix}},
\end{eqnarray}
and consequently, for the Schmidt basis, we write
\begin{eqnarray}\label{Schmidt basis}
\rho_{s}=\det\frac{1}{2}(\mathbb{I}-\textbf{G})[{\rm \det}(\textbf{F}_{s})]^{\frac{1}{2}}e^{\frac{1}{2}(\boldsymbol{\delta}^{\dagger}\,\,\boldsymbol{\delta})\textbf{D}
  \begin{pmatrix}
  \boldsymbol{\delta}\\
\boldsymbol{\delta}^{\dagger}\\
  \end{pmatrix}}.
\end{eqnarray}
Obviously, the above two density matrices are not the same density matrices and one may find their distance by looking at the Frobenius norm distance defined as follows:
\begin{equation}\label{Frobenious}
||\rho_{e}-\rho_{s}||_{_F}=(\text{tr}[\rho_{e}-\rho_{s}]^2)^{\frac{1}{2}}.
\end{equation}
This quantity is a measure of the distance between the density matrices written in the two different bases and can be easily calculated for our density matrices using the following formulas
\begin{eqnarray}\label{traces}
\text{tr}\rho_{s}^2=\text{tr}\rho_{e}^2=\det\frac{1}{2}(\mathbb{I}-\textbf{G})[{\rm \det}(\textbf{F}_{s})]^{\frac{1}{2}}\prod_{i=1}^l(2\cosh\lambda^Q_i),\nonumber\\
\text{tr}[\rho_{s}\rho_{e}]=\det\frac{1}{2}(\mathbb{I}-\textbf{G})[{\rm \det}(\textbf{F}_{s})]^{\frac{1}{2}}\prod_{i=1}^l(2\cosh\frac{\lambda^{\tilde{Q}_i}}{2}),\nonumber\\
\end{eqnarray}
where $\lambda^Q_i$'s are the positive eigenvalues of the matrix $\textbf{Q}$ and $\lambda^{\tilde{Q}}_i$'s are the positive eigenvalues of the matrix $\tilde{\textbf{Q}}$ defined as $e^{\tilde{\textbf{Q}}}=e^{\textbf{Q}}e^{\textbf{D}}$.

\begin{figure} [hthp!] 
\includegraphics[width=0.5\textwidth,angle =0]{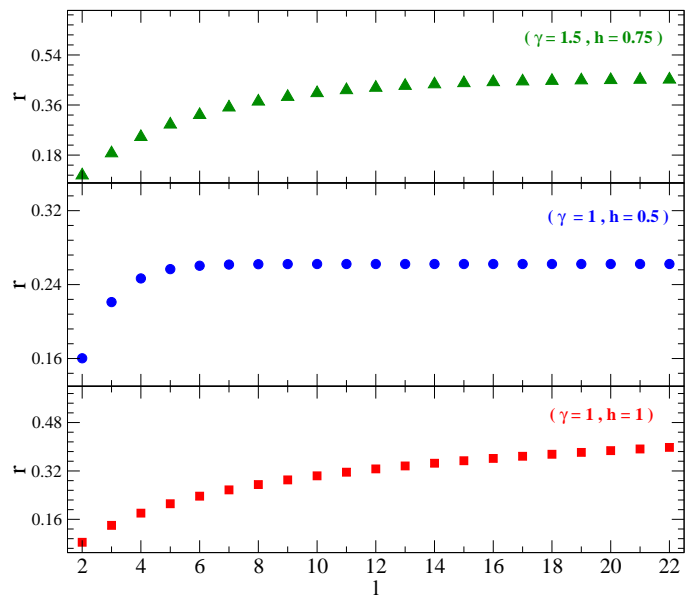}
\centering
\caption{(color online) The norm ratio for different points in the phase diagram of the XY chain. As it is shown in  the different panels, for all of the cases the norm riches a constant value which is smaller than one. For the case of the critical point depicted in the bottom panel, the convergence is slower as one might be expected.  } 
\label{fig_SEnorm}
\end{figure}

To have a measure of how big is the Frobenius distance, we compare it with the Frobenius norm of the density matrix itself and define
\begin{eqnarray}\label{Frobenious}
r=\frac{||\rho_{e}-\rho_{s}||_{_F}}{||\rho_{s}||_{_F}}
\end{eqnarray}
In the Figure \ref{fig_SEnorm}, we depicted this ratio for different points on the phase diagram of the XY chain for different subsystem sizes. The ratio is always smaller than one and saturates for larger subsystem sizes which indicates that the two bases are not far from each other. Note that for two arbitrary density matrices the ratio is normally bigger than one. For example, we realized that this ratio is around one if one shuffles the  order of the eigenvalues of the matrix $\textbf{D}$.

%%%%%%%%%%%%%%%%%%%%%%%%%%%%%%%%%%%%%%%%%%%%%%%%%%%%%
%%%%%%%%%%%%%%%%%%%%%%%%%%%%%%%%%%%%%%%%%%%%%%%%

%%%%%%%%%%%%%%%%%%%
\section{ R\'enyi entropy of the subsystem energy and the Loschmidt echo}
%\label{sec2}
%%%%%%%%%%%%%%%%%%%
It is possible to connect the generating function of our quantity, i.e. $M(z)=\text{tr}[\rho_l e^{zH_l}]$, to the fidelity 
amplitude by first preparing the full system in 
the ground state and then turning off  the couplings between the Hamiltonian of the subsystem with the rest of the system and also, 
switching off all the couplings 
between the spins outside of the subsystem. The current Hamiltonian is now $H_l\otimes I_{\bar{l}}$. 
The we can calculate the Loschmidt amplitude in 
the Schmidt basis, 
i.e. $|\psi_g\rangle=\sum_j\lambda_j|\phi^{(l)}_j\rangle\otimes|\phi^{(\bar{l})}_j\rangle$ 

%\begin{widetext}
\begin{eqnarray}\label{Loschmidt in Schmidt basis}
\langle\psi_g|e^{-it(H_l\otimes I_{\bar{l}})}|\psi_g\rangle&=&\sum_{j,k}\lambda_j\lambda_k
\langle\phi^{(l)}_j|e^{-itH_l}|\phi_k^{(l)}\rangle\delta_{jk}\nonumber\\
&=&\sum_{j}\lambda_j^2
\langle\phi^{(l)}_j|e^{-itH_l}|\phi_j^{(l)}\rangle
\end{eqnarray} 
%\end{widetext}
%
This quantity is nothing except $\text{tr}[\rho_l e^{-itH_l}]$ which leads to
\begin{eqnarray}\label{LE and GF}
M(-it)=\langle\psi_g|e^{-it(H_l\otimes I_{\bar{l}})}|\psi_g\rangle
\end{eqnarray}  
Using the above equation, we end up with a remarkable result that one can generate 
the probability distribution of finding the system in different energy states by just performing an inverse Fourier transform. In other words
\begin{eqnarray}\label{LE and GF2}
P(E_j)=\frac{1}{2\pi}\int dt e^{itE_j}M(-it)
\end{eqnarray}  
Using the above probabilities one can calculate the desired entropies. Notice that the right hand 
side in equation (\ref{LE and GF2}) is the Loschmidt amplitude and remarkably, the inverse Fourier transform of this amplitude gives us the desired probability distributions. 
Finally, there is a number of different experiments which make the measurement of the Loschmidt amplitude possible. In a recent experiment, the time resolved state tomography 
provided a full access to the evolution of the wavefunction in ultracold atoms~\cite{Flaschner2016,Flaschner2018}. In another directly related experiment, an  efficient MPS 
tomography for the XY model has been used to reconstruct the full density matrix for the short range interaction which enables one to measure the Loschmidt amplitude~\cite{Roos2017,Cramer2010}. 
Furthermore, there is relatively an easier method to measure $H_2$ as it is related to the Loschmidt probabilities which has been measured in a recent experiment of 1D 
trapped ions~\cite{Jurcevic2017}. We note that our proposal is reminiscent of the recent measurement quench protocol\cite{Jurcevic2015,Bayat2018}

%%%%%%%%%%%%%%%%%%%
\section{Conclusions}
%\label{sec2}
%%%%%%%%%%%%%%%%%%%

In this paper, we showed that the subsystem energy entropy is an excellent quantity which can mimic the entanglement entropy of many body systems
at and off the critical point. For the non-critical systems,  the defined R\'enyi entropy for sufficiently large $\alpha$'s follows an area law and at the critical point it follows a 
logarithmic behavior with a universal coefficient. It can be used to detect the phase transition and determine the universality class.  We also proposed an experimental 
setup to measure our quantity with the current technology. Since the probabilities introduced in this paper follow closely the Schmidt coefficients our protocol 
can be used to measure the Schmidt coefficients effectively. For example, for the gapped systems the few largest probabilities are extreemly close
to the  biggest Schmidt coefficients.
In another words finding the first largest probabilities provide an approximation for the most relevant Schmidt coefficients.
It would be very important to calculate this quantity in QFT (CFT) to have a better idea about the nature of the universality of the presented results \cite{bosonic}.
The recent developments regarding the distribution of the energy-momentum tensor in QFT(CFT) might be very useful, see Ref. \cite{Fewster2018} and references therein.

\vspace*{0.5cm}

\textbf{Acknowledgement}
 We thank P Calabrese, M. Collura and N. M. Linke
for discussions. We thank E. Barnes  for reading the manuscript and his useful comments. The work of MAR was supported in part by CNPq. The work of K.N. is supported by National
Science Foundation under Grant No. PHY-1620555 and DOE grant DE-SC0018326.

%\appendix
\appendix

\section{Diagonalization of the Free Fermions}
\label{appA}
In this subsection, we summarize the result of Ref.\cite{LSM}. Consider a generic (real) truncated 
free fermion Hamiltonian defined in domain $D$: 
\begin{eqnarray}\label{Hsub}\
\textbf{H}_{D}=
\sum_{ij}[c_{i}^{\dagger}A_{ij}c_{j}+\frac{1}{2}c_{i}^{\dagger}B_{ij}c_{j}^{\dagger}+\frac{1}{2}c_{i}B_{ji}c_{j}]-\frac{1}{2}{\rm Tr}{\textbf{A}},\hspace{1cm}
\end{eqnarray} 
%
%where Tr A coming from?
where $c_{i}^{\dagger}$ and $c_{i}$ are fermionic creation and annihilation operators and $i=1,2,...,|D|$, where $|D|$
is the number of sites in the region $D$.  The Hermitian Hamiltonian requires $\textbf{A}$ and $\textbf{B}$ to be  symmetric and antisymmetric matrices respectively. 
To diagonalize the Hamiltonian we use the following  canonical transformation
\begin{eqnarray}\label{mat_U1}\
\begin{pmatrix}
\textbf{c} \\
\textbf{c}^{\dagger} \\
\end{pmatrix}=
\textbf{U}^{\dagger}
\begin{pmatrix}
\boldsymbol{\eta} \\
\boldsymbol{\eta}^{\dagger} \\
\end{pmatrix},
\end{eqnarray} 
with 
\begin{eqnarray}\label{mat_U2}\
\textbf{U}= \begin{pmatrix}
\textbf{g} & \textbf{h}\\
\textbf{h}^{*} & \textbf{g}^{*}\\
  \end{pmatrix}.
\end{eqnarray} 
Then, we can write the diagonalized from of the Hamiltonian  as follows
\begin{eqnarray}\label{Hdiag}\
\textbf{H}_{D}=\sum_{k}|\lambda_{k}|(\eta_{k}^{\dagger}\eta_{k}-\frac{1}{2}).
\end{eqnarray} 
Note that $\textbf{g}$ and $\textbf{h}$ can be derived from the following equations:
\begin{eqnarray}\label{gandh}\
\textbf{g}&=&\frac{1}{2}(\boldsymbol{\phi}+\boldsymbol{\psi}),\\
\textbf{h}&=&\frac{1}{2}(\boldsymbol{\phi}-\boldsymbol{\psi}),
\end{eqnarray} 
where we have
\begin{eqnarray}\label{phi and psi 1a}\
(\textbf{A}+\textbf{B})\phi_k&=&|\lambda_k|\psi_k,\\
\label{phi and psi 1b}
(\textbf{A}-\textbf{B})\psi_k&=&|\lambda_k|\phi_k,
\end{eqnarray} 
or
\begin{eqnarray}\label{phi and psi 2a}\
(\textbf{A}-\textbf{B})(\textbf{A}+\textbf{B})\phi_k&=&|\lambda_k|^2\phi_k,\\
\label{phi and psi 2b}
(\textbf{A}+\textbf{B})(\textbf{A}-\textbf{B})\psi_k&=&|\lambda_k|^2\psi_k.
\end{eqnarray} 
When $\lambda_k\neq0$, $\phi_k$ and $\lambda_k$ can be calculated by solving the eigenvalue equation (\ref{phi and psi 2a}),
then, $\psi_k$ can be determined using (\ref{phi and psi 1a}). When $\lambda_k=0$, $\phi_k$ and $\psi_k$ 
can be deduced directly from (\ref{phi and psi 1a}) and (\ref{phi and psi 1b}).

The correlation matrix $\textbf{G}$ for the full system defined as 
\begin{eqnarray}\label{Gmat2}
G_{ij}&=&\langle (c_{i}^{\dagger}-c_{i})(c_{j}^{\dagger}+c_{j})\rangle
\end{eqnarray}
can be also calculated using the above procedure as follows:
\begin{eqnarray}\label{Gmat7}
\textbf{G}=(\hat{\textbf{h}}^{\dagger}-\hat{\textbf{g}}^{\dagger})(\hat{\textbf{g}}+\hat{\textbf{h}}).
\end{eqnarray}
Note that in the above equation we put hat on the $\textbf{g}$ and $\textbf{h}$ matrices to emphasize that one should calculate them using the $\hat{\textbf{A}}$ and $\hat{\textbf{B}}$ matrices.
Since we never calculate the correlation matrix for the truncated Hamiltonian we do not put hat on this matrix.

%%%%%%%%%%%%%%%%%%%%%%%%%%%%%%%%%%%%%%%%%%%%%%%%%%%%%%%%%%
\section{Generating function of the subsystem energy}
\label{appB}
%%%%%%%%%%%%%%%%%%%%%%%%%%%%%%%%%%%%%%%%%%%%%%%%%%%%%%%%%%
It is worth mentioning that the results of this section can be used to get also an explicit formula for the generating
function of the subsystem energy. Different versions of this quantity are already appeared in \cite{Najafi2017}. We present here another version
which has different form but it is equivalent to the previous ones. The generating function is defined as:

%\begin{widetext}
\begin{eqnarray}\label{generating function Definition}
M(z)=\text{tr}[\rho_De^{z\textbf{H}_{D}}].
\end{eqnarray}
After writing $\textbf{H}_{D}$ as
\begin{eqnarray}\label{HD}
\textbf{H}_{D}=\frac{1}{2}(\textbf{c}^{\dagger}\,\,\textbf{c})\begin{pmatrix}
\textbf{A} & \textbf{B}\\
-\textbf{B} & -\textbf{A}\\
  \end{pmatrix}
  \begin{pmatrix}
\textbf{c}\\
\textbf{c}^{\dagger}\\
  \end{pmatrix}
\end{eqnarray}
%\end{widetext}
%
and using (\ref{entanglement Hamiltonian}) the trace can be calculated explicitly. The final result is 
\begin{widetext}
\begin{eqnarray}\label{generating function }
M(z)=\det\frac{1}{2}(\mathbb{I}-\textbf{G})[{\rm \det}(\textbf{F}_{s})]^{\frac{1}{2}}\det[\mathbb{I}+e^{z\begin{pmatrix}
\textbf{A} & \textbf{B}\\
-\textbf{B} & -\textbf{A}\\
  \end{pmatrix}}e^{\begin{pmatrix}
\textbf{M} & \textbf{N}\\
-\textbf{N} & -\textbf{M}\\
  \end{pmatrix}}]^{\frac{1}{2}}.
\end{eqnarray}
\end{widetext}
The above equation can be also written as
\begin{eqnarray}\label{generating function 2}
M(z)=&&\det\frac{1}{2}(\mathbb{I}-\textbf{G})[{\rm \det}(\textbf{F}_{s})]^{\frac{1}{2}}\nonumber \\
&&\times\det[\mathbb{I}+\begin{pmatrix}
e^{z|\boldsymbol{\lambda}|} & 0\\
0 & e^{-z|\boldsymbol{\lambda}|}\\
  \end{pmatrix}\tilde{\textbf{T}}]^{\frac{1}{2}};\hspace{1cm}
\end{eqnarray}
where $|\boldsymbol{\lambda}|$ is the matrix of the eigenvalues of the subsystem Hamiltonian. Since the involved matrices normally does not have simple properties it seems difficult to find the analytical properties of the above formula for large subsystem sizes.

%%%%%%%%%%%%%%%%%%%%%%%%%%%%%%%%%%%%%%%%%%%%%%%%%%%%%
\begin{figure}[htbp!] 
\includegraphics[width=0.5\textwidth,angle =0]{epsinf2.eps}
\centering
\caption{(color online)Up: logarithmic behavior of the R\'enyi entropy ($\alpha\to\infty$) of the subsystem energy of the 
critical XY chain for $\gamma=h=0$ and $\gamma=h=1$. Down: area law of the R\'enyi entropy ($\alpha\to\infty$) for various values of the parameters of the gapped XY chain.} 
\label{Renyiinf}
\end{figure}
%%%%%%%%%%%%%%%%%%%%%%%%%%%%%%%%%%%%%%%%%%%%%%%%%

%%%%%%%%%%%%%%%%%%%%%%%%%%%%%%%%%%%%%
\section{Further Numerical Details}
\label{app3}
%%%%%%%%%%%%%%%%%%%%%%%%%%%%%%%%%%%%%

In this section, we will summarize further numerical results regarding the R\'enyi entropy of the subsystem energy. We will mostly focus on the $\alpha\to\infty$ where we can work with the biggest probability. In this case, we can calculate the entropy for relatively large subsystem sizes with high accuracy. For example, in  Fig. \ref{Renyiinf},  we checked the area law for the gapped phase with much more accuracy. Then, in the table I and II we show the universality of
the coefficient of the logarithm, i. e. $\epsilon(\infty)$, in the critical regime for the critical XY line and the XX line respectively. 
In the same tables one can also see that the coefficient corresponding to  the semi-infinite case $\epsilon_s(\infty)$ is half of the infinite case. 
This is exactly what happens for the entanglement entropy\cite{CC2004}. Note that 
the coefficients of the critical Ising universality is with high accuracy half of the the XX universality. This 
indicates that most probably the coefficients $\epsilon(\infty)$ and $\epsilon_s(\infty)$  are linearly proportional to the central charge of the underlying CFT.

\begin{table}[htbp!]
\centering
{\begin{tabular}{|l|c|c|c|}
  \hline
  $(h,\gamma)$  & $\epsilon(\infty)$ & $\epsilon_s(\infty)$\\
\hline
$(1,1)$   & $0.111(1)$  & $0.055(1)$\\
\hline
$(1,0.5)$   & $0.111(1)$  &  $0.055(1)$\\
\hline
$(1,0.75)$  & $0.111(1)$   &  $0.055(1)$\\
\hline
$(1,1.25)$  & $0.111(1)$   & $0.055(1)$\\
\hline
$(1,1.5)$  & $0.111(1)$    & $0.055(1)$\\
\hline
\end{tabular}}
\label{table1}
\caption{The coefficients $\epsilon(\infty)$ and $\epsilon_s(\infty)$ for various  $\gamma$'s on the critical XY line.} 
\end{table} 

\begin{table}[htbp!]
\centering
{\begin{tabular}{|l|c|c|c|}
  \hline
  $n_{c}$ &$ \epsilon(\infty)$ & $\epsilon_{s}(\infty)$ \\
\hline
$\pi/2 $    & $0.222(2)$  & $0.111(2)$ \\
\hline
$14\pi/30$ & $0.222(2)$  & $0.111(2)$\\
\hline
$9\pi/20$  & $0.222(2)$  & $0.111(2)$\\
\hline
$8\pi/18$  & $0.222(2)$  & $0.111(2)$\\
\hline
$7\pi/16$  & $0.222(2)$  & $0.111(2)$ \\
\hline
$6\pi/14$  & $0.222(2)$  & $0.111(2)$\\
\hline
$5\pi/12$  & $0.222(2)$  & $0.111(2)$\\
\hline
$4\pi/10$  & $0.223(2)$  & $0.111(2)$\\
\hline
$3\pi/8$  & $0.224(2)$    &$0.112(2)$\\
\hline
$\pi/3$   & $0.225(2)$   & $0.112(2)$\\
\hline
\end{tabular}}
\label{table1}
\caption{The coefficients $\epsilon(\infty)$ and $\epsilon_s(\infty)$ for various values of the fillings $h=-2\cos n_c$ on the critical XX 
line, i.e. $\gamma=0$. } 
\end{table} 

\newpage
%\section{Reduced Density Matrix in the Local Energy Basis}
%\label{app1}

%\end{document}


\begin{thebibliography}{44}
\bibitem{footnotevolumelaw} The volume law simply reflects that the state spreads over exponential number of configurations.
\bibitem{footnote1} The generalization to short-range interacting systems, finite systems and also higher dimensions is trivial.
\bibitem{footnote2} The precise statement of the theorom is the following: consider an arbitrary eigenstate of the total Hamiltonian  $|E\rangle$
and a sub-region $D$ with complement $\bar{D}$. Define $D'=D\cup D_{b}$ where $D_{b}$ is a region around $D$ which the boundary terms of the sites inside $D$ interact. Then, the theorem tells us that there is a density matrix $\rho_E$ in $D'$ which is weakly diagonal in the eigenbasis of $H_{D'}$ and satisfies $\text{tr}_{\bar{D}}|E\rangle\langle E|\approx \text{tr}_{D_b}\rho_E$. Here, the weekly diagonal means that for two different eigenvectors $|e_{1(2)}\rangle$  of the Hamiltonian $H_{D'}$, we have $|\langle e_1|\rho_E|e_2\rangle|\leq e^{(l-r)(e_1-e_2)^2/8cv^2}$, where $c,v,l$ and $r$ are constants.
\bibitem{footnotealphac}Unfortunately we realized that it is difficlut to calculate $\alpha_c$ effectively. For example, although $\alpha_c$ is for sure smaller than two with 
the current limitations in our numerical
calculations, it is not clear our conclusion can be applied to also the Shannon entropy for the generic values of $\gamma$ and $h$.
\bibitem{footnotesmallP} For example, we checked that the smallest probability which corresponds to the highest energy decays supper-exponentially with the subsystem size, i.e. $\ln p_{\text{min}}\propto e^{-l}$. We believe that the same is true for majority of states with high energies which makes their contribution to the Shannon entropy negligible. One might also understand the absence of the volume-law based on the same line of argument.
\bibitem{footnotesmall}Here we truncate the number of probabilities at $k$ when $p_{k+1}$($\lambda_{k+1}$) is less than $10$ percent of $p_k$($\lambda_k$).
\bibitem{footnotetruncatedRenyi}Similarly one can also define a truncated R\'enyi entropy. Our conclusions are also valid in this case too.
\bibitem{footnoteothermodels}We believe that effectiveness of just taking one(two) probabilities is a property of the XY chain. In other models the number of needed probabilities 
might vary depending on the symmetries of the system but it will not scale with the size of the subsystem.
\bibitem{bosonic} The bosonic systems, especially coupled harmonic oscillators will be studied in an other work.
\end{thebibliography}
\end{document}